\documentclass{jnmp}

\usepackage{amsmath}
\newcommand{\eqdef}{\stackrel{\text{def}}{=}}
\newcommand{\bm}{\boldsymbol}
\newcommand{\sfrac}[2]{{\textstyle \frac{#1}{#2}}}

\JNMPnumberwithin{equation}{section}

\theoremstyle{definition}

\begin{document}

%
\renewcommand{\evenhead}{Ryu Sasaki}
\renewcommand{\oddhead}{New QES Difference Equation}

%
\thispagestyle{empty}

\FirstPageHead{*}{*}{20**}{\pageref{firstpage}--\pageref{lastpage}}{Article}

\copyrightnote{200*}{Ryu Sasaki}

\Name{New Quasi Exactly Solvable Difference Equation}

\label{firstpage}

\Author{Ryu Sasaki}

\Address{Yukawa Institute for Theoretical Physics,\\
      Kyoto University, Kyoto 606-8502, Japan\\
      ~~E-mail: ryu@yukawa.kyoto-u.ac.jp}

\Date{Received Month *, 200*; Revised Month *, 200*; 
Accepted Month *, 200*}

\begin{abstract}
\noindent
Exact solvability of two typical examples of the discrete quantum mechanics, {\em i.e.}
the dynamics of the Meixner-Pollaczek and the continuous Hahn polynomials with {\em full parameters\/}, is newly demonstrated both at the Schr\"odinger and Heisenberg picture levels. A new quasi exactly solvable difference equation is constructed by crossing these two dynamics, that is, the quadratic potential function of the continuous
Hahn polynomial is multiplied by the constant phase factor of the Meixner-Pollaczek type.
Its ordinary quantum mechanical counterpart, if exists, does not seem to be known.
\end{abstract}

\section{Introduction}
As shown recently, Quasi Exact Solvability (QES) is very closely related to
exact solvability \cite{deltaqes,os10,st1}.
If all the eigenvalues of a quantum mechanical system are known together with the corresponding eigenfunctions, the system is exactly solvable in the Schr\"odinger
picture. In contrast, a system is QES if only a finite number (usually the lowest lying ones)
of exact eigenvalues and eigenfunctions are known \cite{ushturb,ush,turb}.
Among various characterisation of quasi exact solvability \cite{ushturb,ush,turb,morozov,gomez}, 
the existence of the invariant polynomial subspace is conceptually simple.
The method to obtain  a QES system, advocated  by the present author \cite{deltaqes,os10,st1}, by deforming an exactly solvable system with an addition/multiplication of a higher order interaction term together with a compensation term,
exemplifies the structure of the invariant polynomial subspace rather clearly
through the action of the similarity transformed Hamiltonian $\tilde{\mathcal H}$ (\ref{tilH3})--(\ref{hact3})
in terms of the pseudo groundstate wavefunction $\phi_0$. This method was applied to the exactly
solvable ordinary quantum mechanics \cite{st1} of one degree of freedom  and multi-particle system
 of Calogero-Sutherland type \cite{cal,sut}. Recently new QES difference equations of one degree of freedom \cite{deltaqes} and multi-particle systems \cite{os10} are obtained by
the application of the same method to the discrete quantum mechanics \cite{os4os5} 
for the Askey-scheme of hypergeometric orthogonal polynomials \cite{And-Ask-Roy,koeswart}
and for the Ruijsenaars-Schneider-van Diejen systems \cite{RS,vD}.

Two of the exactly solvable discrete quantum mechanics discussed in \cite{os4os5,os7,deltaqes},
the Meixner-Pollaczek and the continuous Hahn polynomials,
are of special types in the sense that their parameters are a subset of the allowed ones.

The purpose of the present paper is three-fold. Firstly, to demonstrate the exact solvability of the full dynamics of the 
Meixner-Pollaczek and the continuous Hahn polynomials in the Schr\"odinger picture through shape invariance \cite{genden,os4os5}. The exact Heisenberg operator solutions are also constructed through the closure relations (\ref{closurerel}), (\ref{mpclosure}), (\ref{chclosure}).
The structure of the invariant polynomial subspace is shown explicitly by the action of $\tilde{\mathcal H}$ for each degree monomial (\ref{mpmonoact}), (\ref{chmonoact}).
Secondly, to obtain a new QES difference equation by crossing the above mentioned exactly solvable dynamics. The new system has the quadratic potential with two complex parameters
(\ref{chpot}) coming from the continuous Hahn polynomial and a constant multiplicative phase factor $e^{-i\beta}$ (\ref{mppot}) coming from the Meixner-Pollaczek polynomial.
Thirdly, to give comments on exact Heisenberg operator solutions.
The third part is closely related to the presentation in NEEDS 2007 Workshop by the
present author, ``Heisenberg operator solutions for the Calogero systems" \cite{os9}.

This paper is organised as follows. In section two the exact solvability of the full dynamics 
of the  Meixner-Pollaczek and  continuous Hahn polynomials is demonstrated
after brief review of the general setting of the discrete quantum mechanics appropriate for
the Askey-scheme of hypergeometric orthogonal polynomials. Section three is devoted to the
new  QES difference equation obtained by crossing the dynamics of the full Meixner-Pollaczek and  continuous Hahn polynomials. 
Section four is for the comments on the exact Heisenberg operator solutions. Their dynamical roles,  algebraic  interpretation and 
the connection to a `{\em quantum Liouville theorem\/}' are explained.

\section{Hamiltonian Formulation for  Dynamics of Hypergeometric Orthogonal Polynomials}
It is well known that the classical orthogonal polynomials, the Hermite,
 Laguerre and Jacobi polynomials with various degenerations (Gegenbauer, Legendre, etc) constitute the eigenfunctions of exactly solvable quantum mechanics, for example, the harmonic oscillator without/with the centrifugal potential, the P\"oschl-Teller potential etc.
 Thus it is quite natural to expect that the Askey-scheme of hypergeometric orthogonal polynomials together with their $q$-analogues, which are generalisation/deformation of the classical orthogonal polynomials, also constitute the eigenfunctions of certain quantum mechanics-like systems, so that the orthogonality has proper explanation/interpretation.
In `discrete' quantum mechanics \cite{os4os5}, a Hamiltonian formulation was introduced for the dynamics of several typical examples of the Askey-scheme of hypergeometric orthogonal polynomials. 
Since these polynomials obey  difference equations instead of differential equations, the Hamiltonians contain the
momentum operators in the exponentiated forms in contrast to the second order polynomials in ordinary quantum mechanics.
These examples of discrete quantum mechanics are exactly solvable in the Schr\"odinger picture due to the shape invariance properties \cite{genden,os4os5} and their
exact Heisenberg solutions are given in \cite{os7}. 

In this section we discuss two examples, the Meixner-Pollaczek polynomial and the 
continuous Hahn polynomial, in their full generality. In our previous work on discrete quantum mechanics \cite{os4os5,os7}, only the special
case of the Meixner-Pollaczek polynomial with the phase angle $\phi={\pi}/{2}$
and the special case of the continuous Hahn polynomial with two real parameters $a_1$ and $a_2$ are discussed, partly because these special cases of the two polynomials appear in several other dynamical contexts \cite{degruij,borzov-dama,atakishiyev}
and, in particular, they appear as describing the equilibrium positions \cite{rs1,os2,os4os5} of the classical Ruijsenaars-Schneider van Diejen systems \cite{RS,vD}.

We show that these two polynomials in their full generality, that is, with a general phase angle 
$\phi$ for the Meixner-Pollaczek polynomial and with two complex parameters $a_1$ and $a_2$ for the 
continuous Hahn polynomial, are exactly solvable in the Schr\"odinger as well as in the
Heisenberg picture. Later in section \ref{newqes} we show that a new quasi exactly solvable system is obtained by crossing these  general Meixner-Pollaczek and continuous Hahn polynomials, that is, by multiplying the extra phase factor to the potential of the general continuous Hahn polynomial. The resulting system is no longer exactly solvable but it is quasi exactly solvable by adding a compensation term. 

\subsection{General Setting}
Here we recapitulate the basic notation and rudimentary facts of discrete quantum mechanics of
one degree of freedom. For details we refer to \cite{os4os5,os11}.
The Hamiltonian of a discrete quantum mechanical system of one degree of freedom
to be discussed in this paper has the following general structure
\begin{align}
 \mathcal{H}&\eqdef\sqrt{V(x)}\,e^{ p}\sqrt{V(x)^*}
  +\sqrt{V(x)^*}\,e^{- p}\sqrt{V(x)}-V(x)-V(x)^*
  \label{H}\\
  &=\sqrt{V(x)}\,e^{-i\partial_x}\sqrt{V(x)^*}
  +\sqrt{V(x)^*}\,e^{+i\partial_x}\sqrt{V(x)}-V(x)-V(x)^*,
  \end{align}
in which the potential function $V(x)=V(x\,;\bm{\lambda})$ depends, in general,  on 
a set of parameters $\bm{\lambda}$. 
The exponentiated momentum operators cause a finite shift of the wavefunction in
the {\em imaginary\/} direction: $e^{\pm
i\partial_x}\phi(x)=\phi(x\pm i)$.
As in the supersymmetric quantum mechanics
\cite{infeld, susy}, the
Hamiltonian is always factorised 
\begin{align}
 \mathcal{H}&=A^\dagger A,\\
 A^{\dagger}
  &\eqdef\sqrt{V(x)}\,e^{-\frac{i}{2}\partial_x}
  -\sqrt{V(x)^*}\,e^{\frac{i}{2}\partial_x},\quad
 A \eqdef e^{-\frac{i}{2}\partial_x}\sqrt{V(x)^*}
  -e^{\frac{i}{2}\partial_x}\sqrt{V(x)},
  \label{factor}
 \end{align}
which shows the (formal) hermiticity and positive semi-definiteness of the Hamiltonian.
See the discussion in \S4 of \cite{deltaqes} for detailed realisation of hermiticity.
The groundstate wavefunction $\phi_0(x)$ is annihilated by the $A$ operator
\begin{eqnarray}
  A\phi_0(x)=0 \quad 
  \Longrightarrow  \mathcal{H}\phi_0(x)=0,
  \label{phi0form}
\end{eqnarray}
which can be chosen real $\phi_0(x)\in\mathbb{R}$.
The  eigenfunctions of the Hamiltonian $\phi_n(x)=\phi_n(x\,;\bm{\lambda})$ have the following general structure:
\begin{align}
  \mathcal{H}\phi_n(x)&=\mathcal{E}_n\phi_n(x)\quad
  (n=0,1,2,\ldots),\quad
  0=\mathcal{E}_0<\mathcal{E}_1<\mathcal{E}_2<\cdots,\\
    \phi_n(x\,;\bm{\lambda})
  &=\phi_0(x\,;\bm{\lambda})P_n(\eta(x)\,;\bm{\lambda}).
\end{align}
Here $P_n$ is a polynomial in $\eta(x)$,  which is called a sinusoidal coordinate \cite{os7}.
The orthogonality theorem for the eigenfunctions belonging to different eigenvalues implies that $\{P_n\}$ are orthogonal polynomials with respect to the weightfunction $\phi_0^2(x)$:
\begin{equation}
\int \phi_0^2(x\,;\bm{\lambda})P_n(\eta(x)\,;\bm{\lambda})^*P_m(\eta(x)\,;\bm{\lambda})dx
\propto \delta_{n\,m}.
\end{equation}

\paragraph{Shape Invariance}
If the reversed order Hamiltonian $AA^\dagger$ has the same form as the original 
Hamiltonian $A^\dagger{A}$
\begin{equation}
{A}(\bm{\lambda}){A}(\bm{\lambda})^{\dagger}=
  {A}(\bm{\lambda}+\bm{\delta})^{\dagger}
  {A}(\bm{\lambda}+\bm{\delta})+\mathcal{E}_1(\bm{\lambda}),
  \label{shapeinv}
\end{equation}
the system is called shape invariant \cite{genden,os4os5}. Here $\bm{\delta}$ denotes the shift of the parameters and an additive  constant $\mathcal{E}_1(\bm{\lambda})$ is to be identified
as the energy of the first excited level.
Combined with the basic fact of the supersymmetric quantum mechanics that the
two Hamiltonians $A^\dagger{A}$ and $AA^\dagger$ are iso-spectral (except for the
groundstate), shape invariance determines the entire energy spectrum
and the excited state eigenfunctions from the groundstate wavefunction:
\begin{align}
  &\mathcal{E}_n(\bm{\lambda})=\sum_{s=0}^{n-1}
  \mathcal{E}_1(\bm{\lambda}+s\bm{\delta}),\\
  &\phi_n(x\,;\bm{\lambda})\propto
  {A}(\bm{\lambda})^{\dagger}
 {A}(\bm{\lambda}+\bm{\delta})^{\dagger}
 {A}(\bm{\lambda}+2\bm{\delta})^{\dagger}
  \cdots
 {A}(\bm{\lambda}+(n-1)\bm{\delta})^{\dagger}
  \phi_0(x\,;\bm{\lambda}+n\bm{\delta}).
\end{align}
This establishes the exact solvability in the Schr\"odinger picture.

\paragraph{Heisenberg Operator Solution}
The sinusoidal coordinate $\eta(x)$ has a remarkable property \cite{os7} that the multiple
commutators with the Hamiltonian can be reduced to $\eta(x)$ itself and the 
first commutator $[\mathcal{H},\eta]$ through the closure relation
\begin{equation}
  [\mathcal{H},[\mathcal{H},\eta]\,]=\eta\,R_0(\mathcal{H})
  +[\mathcal{H},\eta]\,R_1(\mathcal{H})+R_{-1}(\mathcal{H}).
  \label{closurerel}
\end{equation}
Here $R_0(\mathcal{H})$ and $R_{-1}(\mathcal{H})$ are in general quadratic polynomials
in $\mathcal{H}$, whereas $R_{1}(\mathcal{H})$ is linear in $\mathcal{H}$. 
This leads to the exact Heisenberg operator solution for the 
sinusoidal coordinate $\eta(x)$:
\begin{align}
  &e^{it\mathcal{H}}\eta(x)e^{-it\mathcal{H}}
  =a^{(+)}e^{i\alpha_+(\mathcal{H})t}+a^{(-)}e^{i\alpha_-(\mathcal{H})t}
  -R_{-1}(\mathcal{H})/R_0(\mathcal{H}),
  \label{H-sol}\\
  &\alpha_{\pm}(\mathcal{H})\eqdef\tfrac12\bigl(R_1(\mathcal{H})
  \pm\sqrt{R_1(\mathcal{H})^2+4R_0(\mathcal{H})}\,\bigr),\\
  &a^{(\pm)}\eqdef\Bigl(\pm[\mathcal{H},\eta(x)]\mp\bigl(\eta(x)
  +R_{-1}(\mathcal{H})/R_0(\mathcal{H})\bigr)\alpha_{\mp}(\mathcal{H})\Bigr)
  \bigm/\bigl(\alpha_+(\mathcal{H})-\alpha_-(\mathcal{H})\bigr).
\end{align}
The entire spectrum $\{\mathcal{E}_n\}$ can also be determined from (\ref{H-sol})
by starting from $\mathcal{E}_0=0$ \cite{os7}, as done by Heisenberg and Pauli
for the harmonic oscillator and the hydrogen atom.
The positive and negative energy parts $a^{(\pm)}$ of the Heisenberg operator solution
$e^{it\mathcal{H}}\eta(x)e^{-it\mathcal{H}}$ are the annihilation-creation operators:
\begin{equation}
  a^{(+)\,\dagger}=a^{(-)},\quad
  a^{(+)}\phi_n(x)\propto \phi_{n+1}(x),\quad
  a^{(-)}\phi_n(x)\propto \phi_{n-1}(x).
\end{equation}

The general theory of exact Heisenberg operator solutions for exactly solvable multi-particle systems is yet to be constructed. For the special case of the Calogero systems \cite{cal,kps}, the totality of Heisenberg operators are derived for any root systems \cite{os9}. For the classical root systems $A$, $BC$ and $D$, the number of particles can be as large as wanted. See section \ref{comments} for comments on exact Heisenberg operator solutions in general.

\subsection{Meixner-Pollaczek polynomial}
\label{MPcase}
The potential function $V(x)$ for the Meixner-Pollaczek polynomial is linear in $x$:
\begin{gather}
  V(x\,;\bm{\lambda})\eqdef e^{-i\beta}(a+ix),\quad  \bm{\lambda}\eqdef a,
  \label{mppot}\\
  \qquad\qquad  0<a\in\mathbb{R},\quad \phi\in\mathbb{R}, \quad \beta\eqdef \phi-\frac{\pi}{2},\quad  0<\phi<\pi.
\end{gather}
The special case discussed in \cite{os4os5,degruij,atakishiyev,borzov-dama} is 
$\beta=0$ or $\phi=\pi/2$.
The groundstate wavefunction $\phi_0$, as annihilated by the operator $A$, $A\phi_0=0$, is given by
\begin{equation}
\phi_0(x\,;a)\eqdef e^{\beta x}|\Gamma(a+ix)|
= e^{(\phi-\frac{\pi}{2})x}\sqrt{\Gamma(a+i x)\Gamma(a-i x)}.
\end{equation}
The similarity transformed Hamiltonian $\tilde{\mathcal H}$ in terms of the 
groundstate wavefunction $\phi_0$,
\begin{align}
 \tilde{\mathcal{H}}\eqdef
  \phi_0^{-1}\circ\mathcal{H}\circ\phi_0
  &=V(x)\left(e^{-i\partial_x}-1\right)+V(x)^*\left(e^{i\partial_x}-1\right)  \label{tilH}
  \\
 &= (a+ix)\,e^{-i\beta}\left(e^{-i\partial_x}-1\right)+(a-ix)\,e^{i\beta}\left(e^{i\partial_x}-1\right)
\label{tilHexpli}
\end{align}
acts on the polynomial part of the wavefunction. 
It is obvious that $\tilde{\mathcal H}$ maps a polynomial into another and it is 
easy to verify
\begin{equation}
 \tilde{\mathcal{H}}\,x^n=2n\cos\beta\, x^n+\mbox{lower order terms}, \quad n\in\mathbb{Z}_+.
 \label{mpmonoact}
\end{equation}
Thus we can find a degree $n$ polynomial eigenfunction $P_n(x)$ of the similarity
transformed Hamiltonian $\tilde{\mathcal{H}}$ 
\begin{equation}
\tilde{\mathcal{H}}P_n(x)=\mathcal{E}_nP_n(x),\quad \mathcal{E}_n=2n\cos\beta
=2n\sin\phi, \quad n=0,1,2,\ldots,
\label{MPeigen}
\end{equation}
which is called the Meixner-Pollaczek polynomial \cite{koeswart}.  It is expressed in terms of the hypergeometric series
\begin{equation}
P^{(a)}_n(x\,;\phi)=\frac{(2a)_n}{n!}e^{in\phi} {}_2F_1\Bigl(\genfrac{}{}{0pt}{}{-n,\,a+ix}{2a}
  \Bigm|1-e^{-2i\phi}\Bigr),
\end{equation}
in which $(b)_n$ is the standard Pochhammer's symbol 
\[
 (b)_n\eqdef\prod_{k=1}^n(b+k-1)=b(b+1)\cdots(b+n-1).
 \]

\medskip
Shape invariance is also easy to verify:
\begin{equation}
{A}(x\,;a){A}(x\,;a)^{\dagger}=
  {A}(x\,;a+\sfrac{1}{2})^{\dagger}
  {A}(x\,;a+\sfrac{1}{2})+\mathcal{E}_1(\bm{\lambda}),\quad 
  \mathcal{E}_1(\bm{\lambda})=2\sin\phi.
  \label{shapeinv2}
\end{equation}
The parameter $a$ is increased by $\sfrac{1}{2}$ whereas the new parameter $\phi$ is invariant. Since $\mathcal{E}_1$ is independent of the shifted parameter $a$, it is trivial to obtain the linear spectrum  $\mathcal{E}_n=2n\sin\phi$, which is the same as (\ref{MPeigen}).

\medskip
The sinusoidal coordinate is $\eta(x)=x$. 
The closure relation (\ref{closurerel}) reads simply
\begin{equation}
[\mathcal{H},[\mathcal{H},x]]=x\,4\sin^2\phi+2\cos\phi\,\mathcal{H}+2a\sin2\phi,
\qquad \alpha_\pm(\mathcal{H})=\pm2\sin\phi,
\label{mpclosure}
\end{equation}
indicating that $x$ undergoes a sinusoidal motion with the frequency $2\sin\phi$.
The Heisenberg operator solution is 
\begin{align}
e^{it\mathcal{H}}\,x\,e^{-it\mathcal{H}}=x\cos[2t\sin\phi]&+i[\mathcal{H},x]\,\frac{\sin[2t\sin\phi]}{2\sin\phi}\nonumber\\
& +\frac{\cos\phi}{2\sin^2\phi}(\mathcal{H}+2a\sin\phi)(\cos[2t\sin\phi]-1).
\end{align}
The annihilation-creation operators are:
\begin{equation}
\,a^{(\pm)}=\pm\,[\mathcal{H},x]/(4\sin\phi)+\frac{1}{2}\left\{x+\frac{\cos\phi}{2\sin^2\phi}(\mathcal{H}+2a\sin\phi)\right\}.
\label{mpan-cr}
\end{equation}
Obviously the expressions (\ref{mpclosure})--(\ref{mpan-cr}) are drastically simplified for the special case of $\phi=\pi/2$ which were discussed in previous work \cite{os4os5,os7}.

\subsection{Continuous Hahn polynomial}
\label{cHahncase}
The potential function $V(x)$ for the continuous Hahn polynomial is quadratic in $x$:
\begin{equation}
  V(x\,;\bm{\lambda})\eqdef(a_1+ix)(a_2+ix), \quad \bm{\lambda}=(a_1,a_2), a_1,a_2\in\mathbb{C}, \quad \mbox{Re}(a_1)>0, \mbox{Re}(a_2)>0.
  \label{chpot}
\end{equation}
The special case discussed in \cite{os4os5,degruij,atakishiyev,borzov-dama} is for real $a_1$ and $a_2$. 
The groundstate wavefunction $\phi_0$, as annihilated by the operator $A$, $A\phi_0=0$, is given by
\begin{equation}
  \phi_0(x\,;\bm{\lambda})\eqdef
  \sqrt{\Gamma(a_1+ix)\Gamma(a_2+ix)\Gamma(a_1^*-ix)\Gamma(a_2^*-ix)}.
\end{equation}
The similarity transformed Hamiltonian $\tilde{\mathcal H}$ in terms of the 
groundstate wavefunction $\phi_0$,
\begin{align}
 \tilde{\mathcal{H}}&\eqdef
  \phi_0^{-1}\circ\mathcal{H}\circ\phi_0
  =V(x)\left(e^{-i\partial_x}-1\right)+V(x)^*\left(e^{i\partial_x}-1\right)  
  \\
 &= (a_1+ix)(a_2+ix)\left(e^{-i\partial_x}-1\right)+(a_1^*-ix)(a_2^*-ix)\left(e^{i\partial_x}-1\right)
\label{tilHexpli2}
\end{align}
acts on the polynomial part of the wavefunction. 
It is obvious that $\tilde{\mathcal H}$ maps a polynomial into another and it is 
easy to verify
\begin{equation}
 \tilde{\mathcal{H}}\,x^n=n(n+a_1+a_1^*+a_2+a_2^*-1) x^n+\mbox{lower order terms}, \quad n\in\mathbb{Z}_+.
 \label{chmonoact}
\end{equation}
Thus we can find a degree $n$ polynomial eigenfunction $P_n(x)$ of the similarity
transformed Hamiltonian $\tilde{\mathcal{H}}$ 
\begin{equation}
\tilde{\mathcal{H}}P_n(x)=\mathcal{E}_nP_n(x),\quad \mathcal{E}_n=n(n+a_1+a_1^*+a_2+a_2^*-1), \quad n=0,1,2,\ldots,
\label{CHeigen}
\end{equation}
which is called the continuous Hahn polynomial \cite{koeswart}.  It is expressed in terms of the hypergeometric series
\begin{gather}
p_n(x\,;a_1,a_2,a_1^*,a_2^*)\\
\qquad=i^n\frac{(a_1+a_1^*)_n(a_1+a_2^*)_n}{n!} {}_3F_2\Bigl(\genfrac{}{}{0pt}{}{-n,n+a_1+a_2+a_1^*+a_2^*-1,\,a_1+ix}{a_1+a_1^*,\, a_1+a_2^*}
  \Bigm|1\Bigr).\nonumber
\end{gather}

\medskip
Shape invariance is also easy to verify:
\begin{gather}
{A}(x\,;a_1,a_2){A}(x\,;a_1,a_2)^{\dagger}=
  {A}(x\,;a_1+\sfrac{1}{2}, a_2+\sfrac{1}{2})^{\dagger}
  {A}(x\,;a_1+\sfrac{1}{2},  a_2+\sfrac{1}{2})+\mathcal{E}_1(a_1,a_2),\\
  \bm{\delta}\eqdef(\sfrac{1}{2},\sfrac{1}{2}),\quad \mathcal{E}_1(a_1,a_2)=b_1,\quad b_1\eqdef a_1+a_2+a_1^*+a_2^*=2\mbox{Re}(a_1+a_2).
  \label{shapeinv3}
\end{gather}
Here we have introduced an abbreviation $b_1$ for convenience.
The parameter $a_1$ and  $a_2$  are increased by $\sfrac{1}{2}$. 
Since $\mathcal{E}_1$ is linearly dependent on the shifted parameters $a_1$,  and  $a_2$, it is trivial to obtain the quadratic spectrum  
$\mathcal{E}_n=n(n+a_1+a_2+a_1^*+a_2^*-1)=n(n+b_1-1)$, which is the same as (\ref{CHeigen}).

\medskip
The sinusoidal coordinate is $\eta(x)=x$. 
The closure relation (\ref{closurerel}) reads simply
\begin{equation}
[\mathcal{H},[\mathcal{H},x]]=x\left(4\mathcal{H}+b_1(b_1-2)\right)+2[\mathcal{H},x]+b_2\mathcal{H}+b_3(b_1-2),
\label{chclosure}
\end{equation}
in which  abbreviations $b_2\eqdef2\mbox{Im}(a_1+a_2)$ and $b_3\eqdef2\mbox{Im}(a_1a_2)$ are used.
The frequencies $\alpha_\pm(\mathcal{H})$ are
\begin{equation}
\alpha_\pm(\mathcal{H})\eqdef1\pm2\sqrt{\mathcal{H}'},\quad
\mathcal{H}'\eqdef \mathcal{H}+(b_1-1)^2/4,\quad 
\mathcal{H}'\phi_n=(n+(b_1-1)/2)^2\phi_n.
\end{equation}
The Heisenberg operator solution reads
\begin{align}
  e^{it\mathcal{H}}xe^{-it\mathcal{H}}
  = & x\,
  \frac{-\alpha_-(\mathcal{H})e^{i\alpha_+(\mathcal{H})t}
  +\alpha_+(\mathcal{H})e^{i\alpha_-(\mathcal{H})t}}
  {4\sqrt{\mathcal{H}'}}+[\mathcal{H}, x]\,
  \frac{e^{i\alpha_+(\mathcal{H})t}-e^{i\alpha_-(\mathcal{H})t}}
  {4\sqrt{\mathcal{H}'}}\nonumber\\[4pt]
  &+\frac{b_2\mathcal{H}+b_3(b_1-2)}{4(\mathcal{H}+b_1(b_1-2))}
  \left(\frac{-\alpha_-(\mathcal{H})e^{i\alpha_+(\mathcal{H})t}
  +\alpha_+(\mathcal{H})e^{i\alpha_-(\mathcal{H})t}} {4\sqrt{\mathcal{H}'}}-1\right).
  \label{quantsol0}
\end{align}
The annihilation and creation operators are:
\begin{align}
  a^{\prime(\pm)}&\eqdef a^{(\pm)}4\sqrt{\mathcal{H}'}\nonumber\\
  &=\pm [\mathcal{H},x]\mp \left(x+\frac{b_2\mathcal{H}+b_3(b_1-2)}{4(\mathcal{H}+b_1(b_1-2))}\right)\,\alpha_{\mp}(\mathcal{H}).
  \label{chan-cr}
\end{align}
Obviously the expressions (\ref{chclosure})--(\ref{chan-cr}) become drastically simple
for the special case of $b_2=b_3=0$, which were discussed in previous works \cite{os4os5,os7}.
\section{New QES Difference Equation}
\label{newqes}
\setcounter{equation}{0}
Here we will discuss the discrete quantum mechanics obtained by crossing the
Meixner-Pollaczek and the continuous Hahn polynomials, that is, with the quadratic potential function of the continuous Hahn polynomial (\ref{chpot}) multiplied by a
constant phase factor $e^{-i\beta}$ of the Meixner-Pollaczek type.
As vaguely expected, the exact solvability is not realised. We will show, instead, that the
system is quasi exactly solvable by adding a compensation term which is linear in $x$:
\begin{align}
 \mathcal{H}&\eqdef\sqrt{V(x)}\,e^{-i\partial_x}\sqrt{V(x)^*}
  +\sqrt{V(x)^*}\,e^{+i\partial_x}\sqrt{V(x)}-V(x)-V(x)^*+\alpha_{\mathcal M}x\\
  &=A^\dagger{A}+\alpha_{\mathcal M}x,\qquad \alpha_{\mathcal M}\eqdef -2\mathcal{M}\sin\beta,\quad \mathcal{M}\in\mathbb{Z}_+,
  \label{factformadded}\\
&\qquad V(x)\eqdef(a_1+ix)(a_2+ix)e^{-i\beta}, \quad a_1,a_2\in\mathbb{C}, 
 \quad  \mbox{Re}(a_1)>0, \mbox{Re}(a_2)>0.
\end{align}
It should be noted that the Hamiltonian is no longer positive semi-definite but the
hermiticity is preserved. The main part, that is without the compensation term, is
factorised as before (\ref{factor}). The zero mode of the $A$ operator
\begin{equation}
A\phi_0=0\Longrightarrow 
  \phi_0(x)\eqdef
  e^{\beta x}\sqrt{\Gamma(a_1+ix)\Gamma(a_2+ix)\Gamma(a_1^*-ix)\Gamma(a_2^*-ix)},
\end{equation}
is no longer the groundstate wavefunction. It is called the pseudo groundstate wavefunction \cite{deltaqes}.

The similarity transformed Hamiltonian $\tilde{\mathcal H}$ in terms of the 
pseudo groundstate wavefunction $\phi_0$,
\begin{align}
 \tilde{\mathcal{H}}&\eqdef
  \phi_0^{-1}\circ\mathcal{H}\circ\phi_0
  =V(x)\left(e^{-i\partial_x}-1\right)+V(x)^*\left(e^{i\partial_x}-1\right)+\alpha_{\mathcal M}x
  \label{tilH3}\\
 &= (a_1+ix)(a_2+ix)e^{-i\beta}\left(e^{-i\partial_x}-1\right)
 +(a_1^*-ix)(a_2^*-ix)e^{i\beta}\left(e^{i\partial_x}-1\right)\nonumber\\
 &\quad -2\mathcal{M}\sin\beta\,x,
\label{tilHexpli2}
\end{align}
acts on the polynomial part of the wavefunction. 
It is obvious that $\tilde{\mathcal H}$ maps a polynomial into another and it is 
easy to verify
\begin{equation}
 \tilde{\mathcal{H}}\,x^n=2(-\mathcal{M}+n)\sin\beta\, x^{n+1}+\mbox{lower order terms}, \quad n\in\mathbb{Z}_+.
 \label{hact3}
\end{equation}
This means that the system is not exactly solvable without the compensation term, but 
it is quasi exactly solvable, since
$\tilde{\mathcal H}$ has an invariant polynomial subspace of degree $\mathcal{M}$:
\begin{align}
\tilde{\mathcal{H}}\,{\mathcal V}_{\mathcal M}&\subseteq
{\mathcal V}_{\mathcal M},
\label{subspc3}\\[4pt]
    {\mathcal V}_{\mathcal M}
&\eqdef
\mbox{Span}\left[1, x, x^2,\ldots,x^{\mathcal 
    M}\right],\quad \mbox{dim}{\mathcal V}_{\mathcal M}={\mathcal M}+1.
    \label{Vdimform3}
\end{align}
 The Hamiltonian ${\mathcal H}$ (\ref{factformadded}) is obviously hermitian
(self-adjoint) and all the eigenvalues are real 
and eigenfunctions can be chosen
real.
 We can obtain a finite number ($\mathcal{M}+1$) of exact eigenvalues and 
eigenfunctions for each given $\mathcal{M}$. The oscillation theorem linking the number 
of eigenvalues (from the groundstate) to the zeros of eigenfunctions
does not hold in the difference equations.
The square integrability of all
the eigenfunctions $\int_{-\infty}^\infty\phi^2(x)dx<\infty$ is obvious.
See \cite{deltaqes} for other examples of quasi exactly solvable difference equations
of one degree of freedom and \cite{os10} of many degrees of freedom.

\medskip
It is easy to demonstrate that multiplying an extra constant phase factor $e^{-i\beta}$
to the other exactly solvable potential functions \cite{os4os5}
\begin{align}
V(x)&\eqdef\frac{(a_1+ix)(a_2+ix)(a_3+ix)}{2ix(2ix+1)},\qquad\qquad\qquad \mbox{continuous dual Hahn},\\
V(x)&\eqdef\frac{(a_1+ix)(a_2+ix)(a_3+ix)(a_4+ix)}{2ix(2ix+1)},\qquad\qquad\qquad \mbox{Wilson},\\
  V(x)&\eqdef\frac{(1-a_1z)(1-a_2z)(1-a_3z)(1-a_4z)}
  {(1-z^2)(1-qz^2)},\quad z=e^{ix},\quad \mbox{Askey-Wilson},
\end{align}
does not provide either exactly solvable or quasi exactly solvable dynamical systems.
The situation is the same for various restrictions of the Askey-Wilson polynomial. 

It is also easy to see that the quasi exact solvability of the systems discussed in \cite{deltaqes} with the potentials
\begin{align}
V(x)&\eqdef{(a_1+ix)(a_2+ix)(a_3+ix)},\\
V(x)&\eqdef{(a_1+ix)(a_2+ix)(a_3+ix)(a_4+ix)},\\
V(x)&\eqdef\frac{(a_1+ix)(a_2+ix)(a_3+ix)(a_4+ix)(a_5+ix)}{2ix(2ix+1)},\\
V(x)&\eqdef\frac{(a_1+ix)(a_2+ix)(a_3+ix)(a_4+ix)(a_5+ix)(a_6+ix)}{2ix(2ix+1)},\\
  V(x)&\eqdef\frac{(1-a_1z)(1-a_2z)(1-a_3z)(1-a_4z)(1-a_5z)}
  {(1-z^2)(1-qz^2)},\quad z=e^{ix},
\end{align}
is destroyed if a constant phase factor  $e^{-i\beta}$
 is multiplied.

\section{Comments on Exact Heisenberg Operator Solutions}
\label{comments}
\setcounter{equation}{0}
Let us start with a rather naive  question; ``What can we learn more
from the exact Heisenberg operator solutions when we know the complete spectrum and the corresponding eigenfunctions?" A small digression on the well known
relationship between the Schr\"odinger and Heisenberg pictures would be useful.
Suppose we have a complete set of solutions of the Shr\"odinger equation
\[
\mathcal{H}\phi_{\bm n}=\mathcal{E}_{\bm n}\phi_{\bm n}.
\]
For {\em any\/} observable $A$, one can construct a (usually  infinite) matrix $\hat{A}$,
$\hat{A}_{{\bm n}{\bm m}}=\langle\phi_{\bm n}|A|\phi_{\bm m}\rangle$, satisfying the Heisenberg equation of motion
\[
\frac{\partial\hat{A}}{\partial t}=i[\mathcal{H},\hat{A}].
\]
Obviously such an exact Heisenberg operator solution does not teach us anything more.

But for a special choice of the observables, called the `sinusoidal coordinates' $\{\eta_j\}$, $j=1,\ldots,r$, with $r$ being the degree of freedom, the operators 
\[
\{e^{i\mathcal{H}t}\eta_j e^{-i\mathcal{H}t}\},\quad j=1,\ldots,r,
\]
can be expressed explicitly in terms of the fundamental operators $\{\eta_j\}$, $\mathcal{H}$ and a finite number of multiple commutators of $\{\eta_j\}$ with the Hamiltonian $[\mathcal{H},[\mathcal{H},[\cdots,\eta_j]..]$. These are the Heisenberg operator solutions found by Odake-Sasaki for a wide class of exactly solvable degree one quantum mechanics
including the discrete ones \cite{os7} and for typical multi-particle dynamics of Calogero type for any root system \cite{os9}. It should be stressed that the existence of 
sinusoidal coordinates is not guaranteed at all. There are several exactly solvable 
degree one quantum mechanical systems for which our construction of the Heisenberg operator solutions does not apply. Various reduced Kepler problems and the Rosen-Morse potentials are the typical examples. See \cite{os7} for more details.
For multi-particle systems, the exact Heisenberg operator solutions are known \cite{os9} 
only for the Calogero systems for any root system \cite{cal,kps}.  There are other well-known exactly solvable multi-particle systems; the Sutherland systems \cite{sut} and the Ruijsenaars-Schneider-van Diejen systems \cite{RS,vD}.
The name `sinusoidal' implies that they all undergo sinusoidal motion but not harmonic.
In classical mechanics terms, the frequencies depend on the initial conditions.

{}From the analysis point of view, the sinusoidal coordinates generate the polynomial eigenfunction $\{P_{\bm n}\}$, $\phi_{\bm n}=\phi_0P_{\bm n}$ ($\phi_0$ is the groundstate wavefunction).
In other words $\{P_{\bm n}\}$  are orthogonal polynomials in $\{\eta_j\}$.
The exact Heisenberg operator solutions for $\{\eta_j\}$ provide the complete set of multi-variable generalisation of the {\em three term recurrence relations\/},
which characterise  orthogonal polynomials in one variable.

As stressed in \cite{os7,os9}, the positive and negative frequency parts of the Heisenberg operator solutions are the sets of {\em annihilation-creation operators\/}.
They generate the entire eigenfunctions algebraically, and thus form a {\em dynamical symmetry algebra\/} together with the Hamiltonian and possibly with the higher conserved quantities (Hamiltonians). The structure of these dynamical symmetry algebras is identified only  for a few special cases of 
degree one, for example, $su(1,1)$.
It is a good challenge to identify the dynamical symmetry algebra and its irreducible representations for each known exact Heisenberg operator solution, single and multi-degree of freedom.
{}From the algebra point of view, the three term recurrence relations for single variable orthogonal polynomials correspond to the Clebsch-Gordan decomposition rules for rank one algebras.
The multi-particle version would simply correspond to the higher rank counterparts of the
Clebsch-Gordan decomposition rules.

{}From a more basic dynamics point of view, one could consider the exact Heisenberg operator solutions and the associated annihilation-creation operators as an explicit but partial realisation of `{\em quantum Liouville theorem\/}'.
 The classical Liouville theorem asserts that one can construct by quadrature only from the complete set of involutive conserved quantities the generating function of 
 a canonical transformation which brings the system to the {\em action-angle\/} form.
In contrast, the usual formulation of quantum Liouville theorem does not say anything about the second half; quantum mechanical counterpart of `bringing to the action-angle form'.
The complete set of the creation-annihilation operators play the corresponding role;
`generating the entire eigenfunctions from the groundstate wavefunction'.
If such generated eigenstates were the simultaneous eigenstates of the complete set of
involutive conserved quantities, one could say that the quantum Liouville theorem is
fully realised.
It seems that there is still some way to go for that goal.

 \section*{Acknowledgements}

R. S. thanks  Satoru Odake for useful comments.
This work is supported in part by Grants-in-Aid for Scientific
Research from the Ministry of Education, Culture, Sports, Science and
Technology, No.18340061 and No.19540179.

\label{lastpage}

\end{document}